# Developmental Constraints on Vertebrate Genome Evolution

Julien Roux[1,2], Marc Robinson-Rechavi[1,2]*

1 Université de Lausanne, Département d'Ecologie et d'Evolution, Quartier Sorge, Lausanne, Switzerland, 2 Swiss Institute of Bioinformatics, Lausanne, Switzerland

**Abstract**

Constraints in embryonic development are thought to bias the direction of evolution by making some changes less likely, and others more likely, depending on their consequences on ontogeny. Here, we characterize the constraints acting on genome evolution in vertebrates. We used gene expression data from two vertebrates: zebrafish, using a microarray experiment spanning 14 stages of development, and mouse, using EST counts for 26 stages of development. We show that, in both species, genes expressed early in development (1) have a more dramatic effect of knock-out or mutation and (2) are more likely to revert to single copy after whole genome duplication, relative to genes expressed late. This supports high constraints on early stages of vertebrate development, making them less open to innovations (gene gain or gene loss). Results are robust to different sources of data—gene expression from microarrays, ESTs, or in situ hybridizations; and mutants from directed KO, transgenic insertions, point mutations, or morpholinos. We determine the pattern of these constraints, which differs from the model used to describe vertebrate morphological conservation ("hourglass" model). While morphological constraints reach a maximum at mid-development (the "phylotypic" stage), genomic constraints appear to decrease in a monotonous manner over developmental time.

**Funding:** We acknowledge funding from Etat de Vaud, Swiss National Science Foundation grant 116798, and the Swiss Institute for Bioinformatics.

**Competing Interests:** The authors have declared that no competing interests exist.

* E-mail: marc.robinson-rechavi@unil.ch

## Introduction

To what extent do the processes of embryonic development constrain genome evolution? Correlations between developmental timing and morphological divergence have long been observed, but the mechanisms and molecular basis of such patterns are poorly understood. The most commonly used measure of selective pressure on the genome, the ratio of non-synonymous to synonymous substitutions ($d_N/d_S$) in protein coding genes, has been of limited help in this case. Stronger constraints have been found on genes expressed in late embryonic stages in Drosophila [1], but most other studies have failed to report robust evidence for a lower $d_N/d_S$ ratio in genes expressed at constrained developmental stages [2–5]. A different approach has been to characterize which genes are duplicated, and which are not: studies of *C. elegans* [2] and Drosophila [6] have found less duplication of genes expressed in early development. These results show that it is possible to identify developmental constraints at the genomic level. They have a few limitations though. One is that the data available has limited the characterization of developmental time to broad categories such as "early" and "late". A second is the difficulty of relating results from two derived invertebrate species, to morphological evolution models in vertebrates [7].

Indeed it is in vertebrates that the fundamental models of developmental constraint on evolution have been established, starting in the nineteenth century with the "laws" of von Baer [8], claiming a progressive divergence of morphological similarities between vertebrate embryos, with the formation of more general characters before species-specific characters. Integration of these observations within evolutionary biology has not always been straight-forward [9–11]. More recently, an "hourglass" model was proposed to describe morphological evolution across development [12,13]: in the earliest stages of development (cleavage, blastula) there is in fact a great variety of forms in vertebrate embryos. Later in development, a "phylotypic" or conserved stage is observed, where many morphological characteristics are shared among vertebrates. This stage is usually presumed to be around the pharyngula stage. After this bottleneck, a "von Baer-like" progressive divergence is again observed. The conserved phylotypic stage has been explained by assuming higher developmental constraints [13–15]. The limits on morphological evolution would be placed by the structure of animal development, making some changes unlikely or impossible. How such limitations are encoded in the genome, or impact its evolution, is still an open question.

In this work, we investigate the existence and timing of constraints on genes expressed in vertebrate development. We use representatives of the two main lineages of vertebrates, a teleost fish and a tetrapode, and we explore the impact of experimental gene loss, and of gain of gene copies in evolution. We find that timing of development has a strong impact in both cases, but that the pattern of constraints on genome evolution does not follow the morphological hourglass model. High constraints are present in early stages of development and relax progressively over time.

## Results

### Constraints on Gene Loss-of-Function in Zebrafish

First, we used the phenotypes of gene loss-of-function as an indicator of selective pressure on genes. We extracted genes

**Author Summary**

Because embryonic development must proceed correctly for an animal to survive, changes in evolution are constrained according to their effects on development. Changes that disrupt development too dramatically are thus rare in evolution. While this has been long observed at the morphological level, it has been more difficult to characterize the impact of such constraints on the genome. In this study, we investigate the effect of gene expression over vertebrate developmental time (from early to late development) on two main features: the gravity of mutation effects (i.e., is removal of the gene lethal?) and the propensity of the gene to remain in double copy after a duplication. We see that both features are consistent, in both zebrafish and mouse, in indicating a strong effect of constraints, which are progressively weaker towards late development, in early development on the genome.

essential for the viability of the zebrafish, giving a lethal phenotype when non functional [16]. We expect that the loss of a gene should be more deleterious if this gene is expressed at a developmental stage with strong constraints. Thus we estimated whether genes were expressed or not at each stage, and computed the ratio of expressed essential genes to expressed reference genes (no reported loss of function phenotype). We then plotted the variation across development of this ratio. We used two different types of data to evaluate the presence of gene expression: (i) expression patterns from in situ hybridizations (Figure 1A), and (ii) "present" or "absent" calls from an Affymetrix microarray experiment (Figure 1B). Results are consistent for both data types: the proportion of essential genes is higher among genes expressed in early development, with a significant negative correlation. For the in situ hybridizations (Figure 1A), a linear regression is significant, but a parabola is not. The parabola has been suggested as the quantitative expectation of an hourglass-like model [3,17]. These results indicate a continuous trend over developmental time, with stronger constraints on early development.

Considering gene expression either "present" or "absent" allows straightforward statistical analysis, but it is a strong approximation of the continuous nature of gene expression. To take advantage of the quantitative signal from the microarray data, we contrasted the median expression level of all the essential genes to that of all of the reference genes (Figure 2A). We used the median because it is less sensitive to extreme values [18]; results were consistent using the mean (not shown). To estimate the significance of the difference between the two curves, we performed a randomization test (see Methods), which provides 1% and 1‰ confidence intervals (Figure 2B). The expectation is now that the essential genes should be enriched in genes highly expressed at the stages with strong constraints. And consistently with the previous observations, essential genes are significantly more expressed in early stages (until 11.7 hours), and less expressed in late stages of development (from 5 days to 14 days). No specific trend is visible around the phylotypic stage. Similar results are obtained for genes which give an "abnormal" phenotype after loss of function (Text S1 and Figure S4).

To complement this approach, we defined groups of genes according to their expression pattern during development (see Methods). This clustering of zebrafish genes provided us notably with a cluster of 2446 genes with high expression in early development, decreasing over time (Figure 3, cluster 3), and an opposite cluster of 1123 genes lowly expressed in early development, increasing over time (Figure 3, cluster 4). As expected, genes whose expression is highest in early development are more frequently essential (1.1% vs. 0.6%), and induce more frequently abnormal phenotypes when non functional (6.1% vs. 2.9%).

### Constraints on Gene Loss-of-Function in Mouse

We performed a similar analysis in mouse, with some differences of methodology due to the data available. For expression, we used of a large amount of EST (Expressed Sequence Tags) data from libraries spanning development, from which we deduced presence or absence of expression (see Methods). Only phenotypes obtained by the targeted knock-out technique were used. As knock-out experiments with no observable phenotype are reported in mouse, we can use these as a reference set, instead of non annotated genes as in zebrafish. The ratio of expressed essential genes to expressed reference genes is significantly negatively correlated with developmental time (Figure 4A), as in zebrafish (Figure 1).

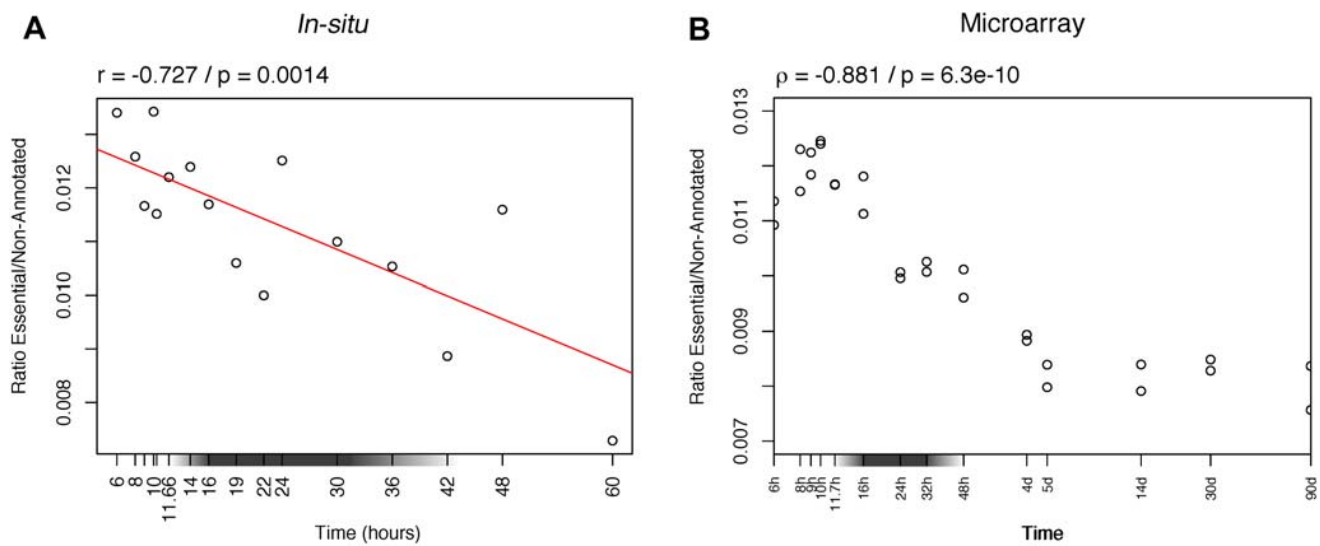

**Figure 1. Variation across zebrafish development of the expression of essential genes compared to non-annotated genes.** At each time point, the ratio of the number of essential genes expressed on the number of non-annotated genes expressed is plotted. A gray box on the x-axis indicates the phylotypic period. (A) Gene expression as reported using in situ hybridization data. The x-axis is proportional to time. A weighted linear regression was fitted to the data and the regression line plotted. (B) Gene expression as reported by "present" calls from Affymetrix array data. The x-axis is in logarithmic scale. A Spearman correlation was computed (coefficient ρ).
doi:10.1371/journal.pgen.1000311.g001

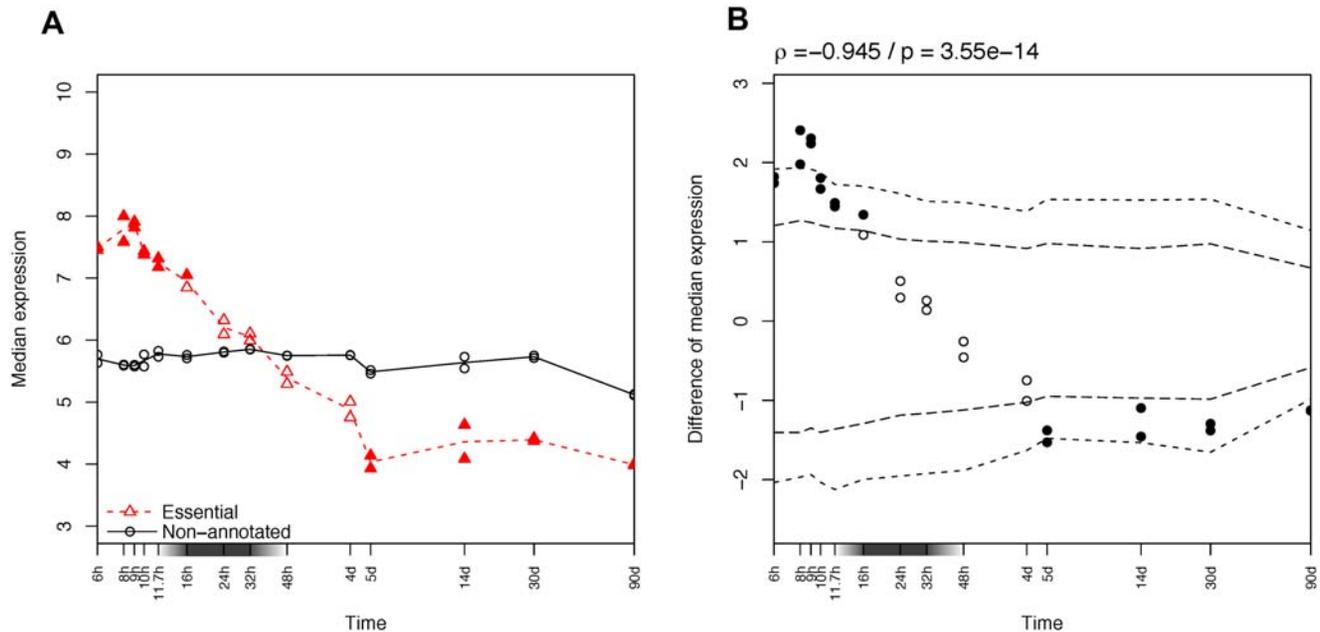

**Figure 2. Expression in zebrafish development of essential genes.** (A) Median expression profiles of zebrafish essential genes, in red dashed line and triangles, compared to non-annotated genes in black solid line and circles. (B) Significance of the expression difference between the two groups of genes. 1‰ and 1% confidence intervals are drawn in dashed lines. Significant points (outside the 1% confidence interval) are filled on both plots. A Spearman correlation was computed (coefficient $\rho$) to test the trend over time. The x-axis is in logarithmic scale. A gray box on the x-axis indicates the phylotypic period.
doi:10.1371/journal.pgen.1000311.g002

Repeating the same approach with genes inducing a phenotype reported as "abnormal" when they are not functional, no significant trend is detected compared to genes inducing no phenotype, after multiple testing correction (Figure 4B). Moreover, these genes can be used as a reference for essential genes (Figure 4C), with results very similar to the use of genes inducing no phenotype after loss of function (Figure 4A). Thus in mouse, genes inducing abnormal phenotypes when non-functional have a behavior more similar to the reference set of "non essential" genes.

## Constraints on Gene Duplication

The fish specific whole genome duplication [19] provides us with a natural experiment on constraints on gene doubling: after this event approximately 85% of duplicated genes lost one copy, and the subset which retained both copies is known to be biased relative to function and selective pressure [20]. Thus we tested if duplicate gene expression pattern in zebrafish development was biased compared to singletons. We plotted the median expression profiles of duplicates originating from the fish specific whole genome duplication, and of singletons, genes whose duplicate copy has been lost after the genome duplication (Figure 5). Duplicates are less expressed in early stages of development. The difference of median expression decreases progressively, similar to the observations for essential or abnormal phenotype genes. Larval time points show a maximum expression of duplicates relative to singletons.

Two scenarios can explain this result. First, retention of two copies may be more likely after the whole genome duplication for genes less expressed in early development. Second, the retention of genes may be unbiased relative to development, but duplicate genes may evolve secondarily lower expression in early development. To get a proxy of the ancestral state before whole genome duplication, we used again mouse data, which has diverged from zebrafish before the fish specific duplication. We compared mouse orthologs of zebrafish duplicates to mouse orthologs of zebrafish singletons, regarding their expression in development (Figure 6). Mouse orthologs of duplicates are significantly less expressed in early development compared to orthologs of singletons. This result in mouse is consistent with the observations in zebrafish, and the most parsimonious explanation is that expression was similar in the ancestor of the two lineages. Therefore we can accept the first hypothesis: after the fish specific whole genome duplication, there was preferential retention of duplicates less expressed in early development.

To check if this phenomenon is particular to the fish specific genome duplication, we repeated this analysis with the two ancient rounds of genome duplication ("2R"), which occurred in the ancestor of vertebrates [21]. It is difficult to distinguish between the two whole genome duplications since no model species diverged from the vertebrate lineage between them. Therefore we looked at the median expression profiles of genes with any duplication at the origin of vertebrates, compared to singletons, whose duplicates were lost after both whole genome duplications. For zebrafish, we restricted this analysis to genes which are singletons regarding the fish specific whole genome duplication. Similarly to fish specific duplicates, duplicates from 2R are significantly less expressed than singletons in the early development of zebrafish (Figure S1) and mouse (Figure S2). Thus mechanisms of retention after whole genome duplication seem to be conserved during vertebrate evolution (see also Text S1).

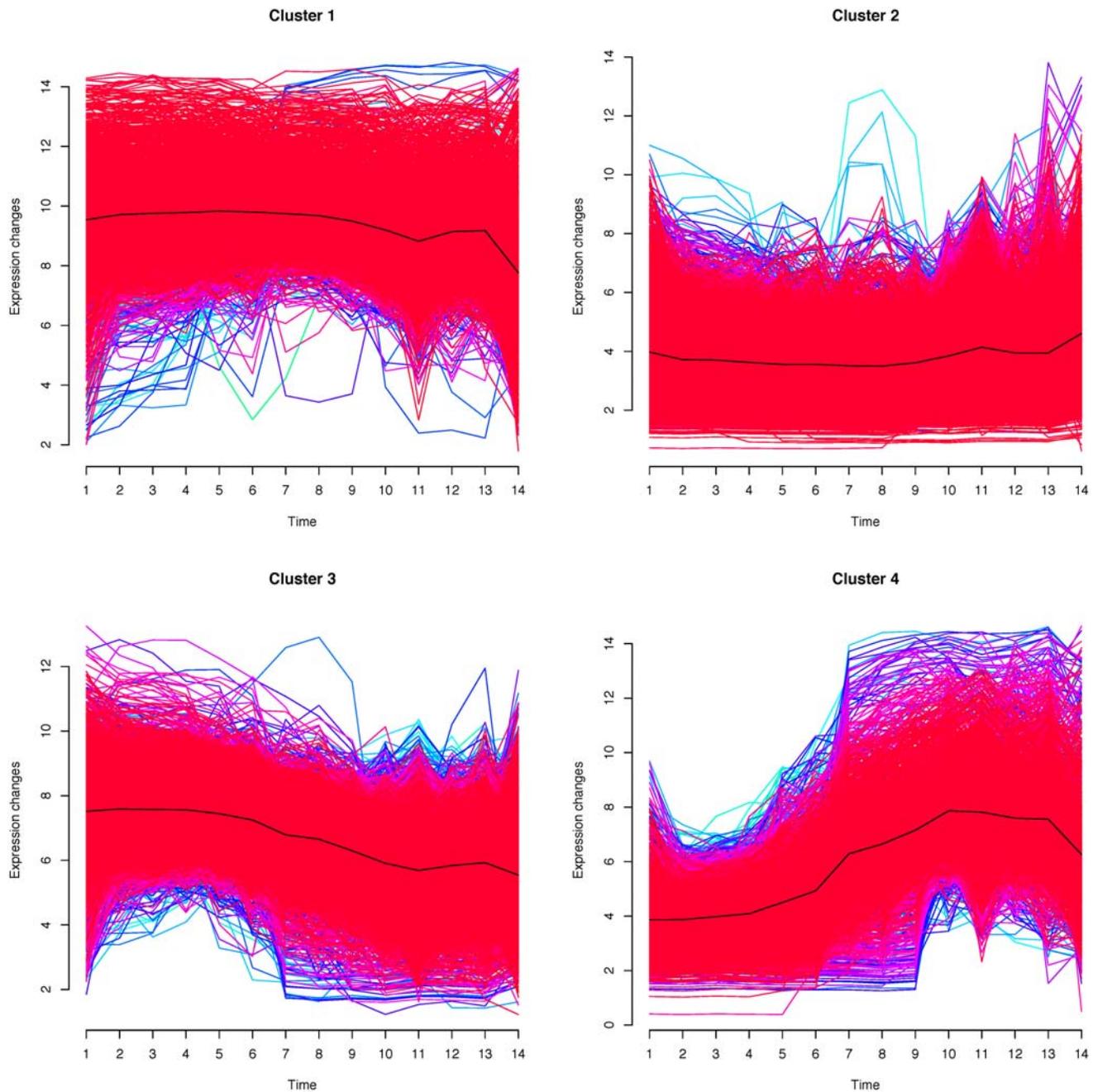

**Figure 3. Expression of four groups of genes, clustered according to their expression in zebrafish development.**
doi:10.1371/journal.pgen.1000311.g003

### Constraints on Gene Sequence

To check if sequences of genes expressed at different stages in development are experiencing different selective pressure, we used the non synonymous to synonymous substitution ratios ($d_N/d_S$). In zebrafish, we used an approach similar to Davis et al. [1]: at each stage we performed the correlation between $d_N/d_S$ and gene expression from microarray data (Figure S3). It has been shown that genes retained in duplicate tend to evolve slowly [20,22]. To control for that factor, we kept only strict singletons in the analysis (genes whose duplicate was lost after 2R and fish-specific genome duplications). At all stages the correlation is negative, confirming that genes with higher expression levels are under stronger purifying selection [23,24]. We note that correlation at the "adult" stage (90 days) is weaker (Figure S3): the link between expression and selective constraints on sequences appears stronger in development than in adult. But there is not a significant trend over time (Spearman $\rho = 0.08$; p = 0.68).

In mouse, we considered only singletons after 2R genome duplication, and we compared the slowest evolving genes (25% lower $d_N/d_S$) with the fastest evolving genes (25% higher $d_N/d_S$). There is a significant correlation with time of expression (Figure 7). Genes with strong sequence constraints (low $d_N/d_S$) tend to be expressed early in development.

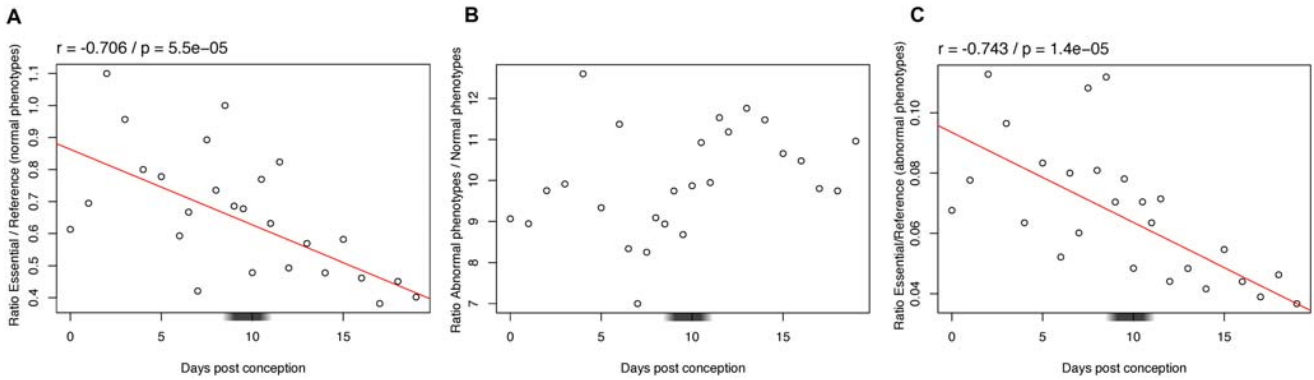

**Figure 4. Variation across mouse development of the ratio of genes with different Knock-Out phenotypes.** (A) Ratio of expressed essential genes relative to "non essential" genes. At each time point, the ratio of the number of essential genes expressed on the number of "non essential" genes expressed is plotted. Detailed counts for each data point in Dataset S2. A weighted linear regression was fitted to the data and the regression line plotted. A Bonferroni multiple-testing correction was used to adjust the significance threshold ($\alpha = 0.05/6 = 0.0083$). A gray box on the x-axis indicates the phylotypic period. (B) Ratio of expressed genes inducing abnormal phenotypes when non functional compared to non essential genes. The linear regression is not significant after multiple testing correction ($r = -0.477$; $p = 0.014$). (C) Ratio of expressed essential genes compared to genes inducing abnormal phenotypes when non functional. Legend as in Figure 4A.
doi:10.1371/journal.pgen.1000311.g004

### Gene Ontology Characterization

What is the function of the genes whose evolution is constrained by expression in early development? We analyzed enrichment or depletion in Gene Ontology [25] categories for the clusters based on gene expression (Figure 3). Using the Molecular Function ontology, genes whose expression is highest in early development are significantly enriched in fundamental processes of the cell, such as RNA processing, transcription, and DNA replication (Table S1). This is very similar to the categories observed to be enriched in house keeping genes [26]. It is also consistent with the categories depleted in fish specific duplicates [20]. Conversely, genes highly expressed in early development are depleted in receptor or channel activity, while these activities are enriched in genes highly expressed in late development. Fewer terms are significant for the Biological Process ontology, and results are essentially consistent with the Molecular Function. Overall, the genes expressed in early development, which appear constrained against gene duplication or loss of function, seem to be house keeping genes involved in basic cellular processes.

### Discussion

Recent discussion of the evolution of ontogeny [27] has allowed the clarification of several important points. The first is that models must be explicitly defined, to allow testing. Poe and Wake [17] distinguish three models for the evolution of ontogeny: the early conservation model *à la* von Baer [8]; the hourglass model, characterized by a conserved phylotypic stage [12,28]; and the adaptive penetrance model (an inverted hourglass). The second point is that quantitative testing is important to distinguish between these models. At the morphological level, several studies have used heterochrony data from vertebrates to quantify the amount of change at each stage of development [17,29]. Surprisingly, this led to rejection of both the early conservation and the hourglass models, although which model is favoured remains disputed [27]. The third point that should be clarified is the distinction between constraints at the level of patterns, and constraints at the level of processes [29]. The studies of heterochrony in vertebrates are typically concerned with the pattern.

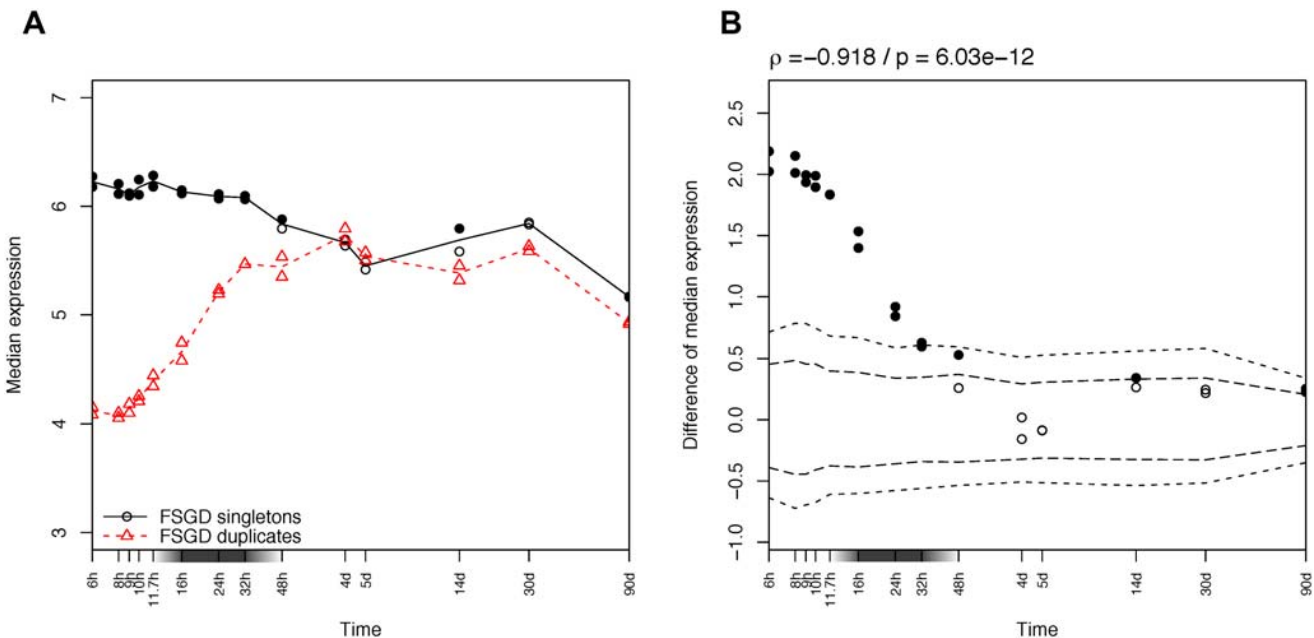

**Figure 5. Expression in zebrafish development of genes according to retention after the fish specific whole genome duplication.** Median expression profiles of zebrafish duplicates from the fish specific whole genome duplication in red dashed line and triangles, and of singletons in black solid line and circles. Legend as in Figure 2.
doi:10.1371/journal.pgen.1000311.g005

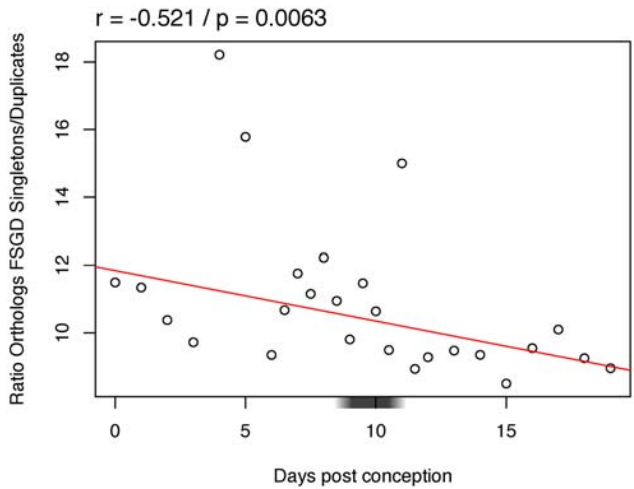

**Figure 6. Variation across mouse development of the ratio of expressed orthologs of zebrafish singletons after the fish specific genome duplication (FSGD) relative to orthologs of zebrafish duplicates.** Legend as in Figure 4.
doi:10.1371/journal.pgen.1000311.g006

In this framework, our results clearly provide a quantitative test which supports the early conservation model. By studying not morphological structures but features of the genome and its expression, this test concerns the level of processes, not patterns. Thus an important point to be made is that our results should be taken neither in contradiction nor in support of any specific model at the level of patterns, given our still limited knowledge of causal relationships between process and patterns in ontogeny [30]. On the other hand, our results do appear to be in contradiction with previous reports of a maximum of constraints on processes around the phylotypic stage of vertebrates [3,4,31].

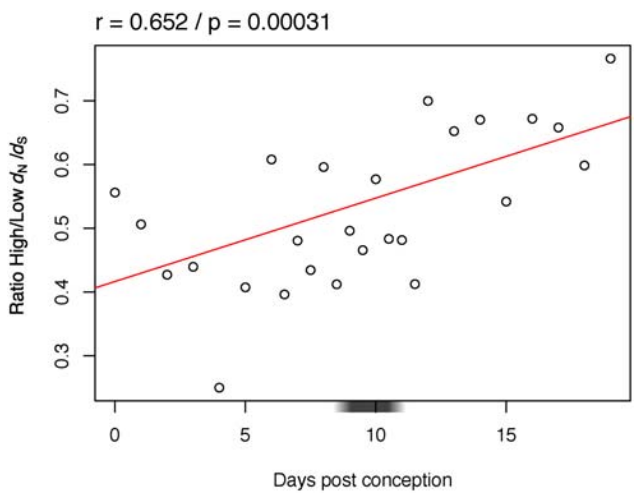

**Figure 7. Variation across mouse development of the expression of rapidly evolving genes (25% highest $d_N/d_S$) compared to slowly evolving genes (25% lowest $d_N/d_S$).** Only singletons for 2R were considered. Legend as in Figure 4.
doi:10.1371/journal.pgen.1000311.g007

We use two simple measures of constraint on the expression of a gene at a developmental stage: if expression of one copy is needed, then (i) removing it may be deleterious, and (ii) increasing the number of copies may also be deleterious. This view is consistent with a recent study in yeast which suggests that constraints influencing the ability to lose certain genes or to maintain them in duplicate may be similar [32]. We expect gain or loss of genes highly expressed at more constrained developmental stages to be counter-selected. And indeed, we find a clear and significant trend: early development is strongly constrained, then constraints diminish during development in a continuous manner. Genes highly expressed in early development are more frequently essential, and less frequently preserved in double copy after genome duplication. Thus early development is less robust against gene loss and against gene doubling. Trends are conserved between mouse and zebrafish, representatives of the two main lineages of bony vertebrates, and between 2R and fish specific genome duplications. An indication of how strong these constraints are is our capacity to predict which genes were kept in duplicate in zebrafish based on expression pattern in mouse. Despite more than 400 MY of independent evolution, and the use of relatively noisy data (mix of EST libraries), more than a quarter of the variance in gene retention is explained (Figure 6; $r^2 = 0.27$). There is also some signal for early conservation at the level of coding sequences, at least in mouse (Figure 7). What we do not see is any genomic evidence for specific constraints at a phylotypic stage. Both in zebrafish and in mouse, the pharyngula stage appears to be part of the general trend from stronger genomic constraints in early development, towards weaker genomic constraints at later stages. We believe that our data are sufficiently detailed, and exhibit sufficiently strong signal, that a maximum of genomic constraints at the phylotypic stage would be visible. So where does the contradiction with previous studies come from?

An early quantitative study [31] found that when screens were done in rodents for the induction of teratogenesis, most abnormalities were obtained by applying teratogens during the phylotypic stage. This was interpreted [31] as supporting strong constraints at the phylotypic stage, due to inductive interactions. But these screens aimed not to test developmental robustness, but to obtain abnormal embryos for experimental work. As remarked by Bininda-Emonds et al. [29], Galis and Metz [31] define the phylotypic stage broadly as including most organogenesis. If application of teratogens in early development resulted in lethality before organogenesis, it would not be of interest to the researchers performing the screens. Thus it seems that what Galis and Metz [31] measured was the potential for a stage to produce morphological abnormalities, not the overall constraints on ontogeny at each stage. There seems to be little reason to suppose that such data provide ''an accurate model of natural selection'' [33], unlike e.g. the retention of duplicate genes over long evolutionary periods.

It is worth noting that we observe a ''peak'' of constraints shortly after pharyngula (Figure 4B) for the expression profile of mouse genes which give an ''abnormal'' phenotype when knocked-out. The behavior of these genes is surprising, because in zebrafish the trend for such genes was similar to that for essential genes. We suspect that the definition of abnormal phenotypes differs between databases and between investigators working in different species. Less severe phenotypes may be reported as "abnormal" in mouse, relative to zebrafish. Of note, data in ZFIN [16] come mainly from the reviewed literature, where minor abnormalities of phenotype are rarely reported, whereas data in the MGD [34] come also from genome wide mutagenesis, and thus include such minor abnormalities. Minor abnormalities in mouse phenotype may also be easier to detect because of the gross similarity with human in anatomy and physiology. In any case, these are the data in our study which most closely approximate the teratogenesis study, and the only data that do not support the early conservation model. Although this trend is statistically not significant, it is consistent with the observations of Galis and Metz [31]. This deserves to be further examined in future studies.

Two other studies which quantified a maximum of constraints at the phylotypic stage did use evolutionary measures of constraint. These studies [3,4] estimated constraints on the evolution of coding sequences, in relation to the timing of expression in mouse development from EST data. Despite similar experimental designs and data, we reached differing conclusions. First, we note that we did check for sequence conservation ($d_N/d_S$) trends over development. In zebrafish, we found no robust pattern (Figure S3), while in mouse we found support for the early conservation model (Figure 7). Second, in our analyses we found that small samples of ESTs could introduce important variability, which is why we used weighted regressions for all computations based on these data. For example, we see a very high ratio of mouse orthologs of zebrafish singletons to duplicates for Theiler stage 5 (day 4) (Figure 6); but this is obtained based on only 628 genes with at least one EST at that stage (median over all stages: 3767). The weighted regression insures that such a point has a weak incidence on the statistical significance. Similar issues are visible in the data of Irie et al. [4], but are not addressed in their analysis. Indeed, the extreme points they use to support constraints at pharyngula are based on some of the smallest samples of their dataset. Finally, it should be noted that another study in mouse found an opposite pattern (relaxation of constraints near the phylotypic stage) using an alternative measure of constraints on sequences, the ratio of radical to conservative amino acid changes, $K_R/K_C$ [5]. In our opinion, these contradictory and weakly supported results are consistent with the idea that overall, coding sequence change seems to have a rather modest contribution to the evolution of development. This is consistent with a stronger contribution of regulation of expression [35,36].

Our results were obtained on data which either reflect the action of natural selection (duplicate gene retention), or are directly relevant to fitness (loss-of-function lethality), and provide unambiguous trends with strong statistical support. Moreover, the consistent patterns in zebrafish in situ hybridization and microarray data, and mouse EST data, show robustness to potential experimental biases or sampling errors. The early conservation model for genomic processes is reinforced by the enrichment of early expressed genes in fundamental cellular processes (Figure 3; Table S1). This is the opposite of duplicated genes, which may be more involved in innovation, and have been reported to be enriched in developmental or behavioural processes [20,21]. Our results are consistent with the observation that basic cores of gene regulatory networks (GRNs) are highly constrained in early stages of animal development [37,38], although we add the notion of a progressive decrease in constraints. This indicates that some relations between the timing of cell-fate decisions in development and rates of genome evolution may be widely shared among animals [7,39]. Indeed, many studies underline gastrulation as a crucial step in development [40,41]. Accordingly this period is shown here to be subject to highest constraints, consistent with the famous Lewis Wolpert quote: "It is not birth, marriage, or death, but gastrulation, which is truly the most important time in your life" [42].

## Materials and Methods

### Microarray Data

Microarray data of zebrafish (*Danio rerio*) development were downloaded from ArrayExpress (E-TABM-33) [43]. This experiment uses an Affymetrix GeneChip Zebrafish Genome Array (A-AFFY-38). 15 stages were sampled, spanning from fertilization to adult stages (15 minutes, 6, 8, 9, 10, 11.7, 16, 24, 30 hours, 2, 4, 5, 14, 30, 90 days, covering zygote, segmentation, gastrula, pharyngula, hatching, larval, juvenile, adult). Two replicates were made per time point; we use both of them for computations, and the 2 values are plotted to give an order of the variability between replicates.

Raw CEL files were renormalized using the package gcRMA [44] of Bioconductor version 2.2 [45]. We used the "affinities" model of gcRMA, which uses mismatch probes as negative control probes to estimate the non-specific binding of probe sequences. The normalized values of expression are in log2 scale, which attenuates the effect of outliers. Mapping of *D. rerio* genes on Affymetrix probesets was made using Ensembl [46] annotation for zebrafish genome version Zv7 (unpublished).

We did not consider the first time point of the data (15 minutes, fertilization). Its behaviour was peculiar in many cases. We explain this by the presence of maternal transcripts in the embryo [47]. These transcripts are largely degraded by 6 hours of development [48], the second time point of the dataset.

For the absolute detection of transcripts (presence or absence calls), the method we used [49] replaces all MM probe values by a threshold value which is based on the mean PM value (after gcRMA transformation) of probesets that are very likely to have absent target transcripts. This removes the influence of probe sequence affinity and results in better performance than the MAS 5 algorithm.

### Significance of Trends in Zebrafish Development

For the zebrafish microarray data we first used a randomization approach to assess the significance of the difference between two curves of median expression across development (for example median expression of duplicates vs. singletons, or of essential genes vs. genes with no reported phenotype). If the two groups contain $n_1$ and $n_2$ genes, we pooled all these genes and randomly separated them into two new groups of same sizes ($n_1$ and $n_2$). Then we calculated and recorded the difference between the two new curves of median expressions across development. After repeating this randomization 10,000 times, we could define 1‰ and 1% confidence intervals.

Second, we calculated the Spearman correlation between developmental time and the difference between two curves of median expression across development. Bonferroni correction was applied to correct for multiple testing, considering the 9 tests computed with this microarray data (Figure 1; Figure 2; Figure 5; Figure S1; Figure S3; Figure S5A–D): $\alpha = 0.05/9 = 0.0056$.

### Clustering of Microarray Data

In order to identify genes lowly or highly expressed in early development, we used the Fuzzy C-Means soft clustering method implemented in the Bioconductor package Mfuzz [50]. After a pre-filtering step (genes with sd <0.5 were removed), we ran the algorithm with the number of clusters set to c = 4. This gave one cluster of genes lowly expressed across development (3641 probesets, 2261 Ensembl genes), one of genes highly expressed (2175 probesets, 1175 Ensembl genes), one of genes whose expression increased (1714 probesets, 1123 Ensembl genes) and one of genes whose expression decreased (3306 probesets, 2446 Ensembl genes) (Figure 3).

### Mouse EST Data

EST (Expressed Sequence Tags) data were retrieved from BGEE (dataBase for Gene Expression Evolution, http://bgee.unil.ch/), a database comparing transcriptome data between species [51], including EST libraries from UniGene [52]. The mapping of UniGene clusters on Ensembl genes is taken from Ensembl (version 48) [46], where a percentage of identity of 90% is set as the minimum threshold to link an Ensembl gene with a UniGene cluster. Each library has been annotated manually to ontologies of anatomy and developmental stages, if it was obtained under non pathological conditions, with no treatment ("normal" gene expression). We considered a gene expressed at one time point in development if at least one EST was mapped to this gene at this time point. Thus, we could retrieve the number of genes expressed at each time point of mouse (Mus musculus) development. From this set we extracted two groups to compare (for example essential/non essential, or duplicates/singletons). As the total number of ESTs available at each time point is different, we use at each time point the ratio of the numbers of genes expressed in the two groups. We obtained similar results when we defined a gene as expressed if it had at least two ESTs mapped to it. Also, considering the ratio of the mean number of ESTs per gene at each stage, instead of the ratio of the number of genes expressed at each stage, gave similar results (not shown). We used data from 297 EST libraries, spanning 26 different developmental stages (from TS01 to TS27), corresponding to a total of 633,307 ESTs.

A weighted linear regression between developmental time and expression ratios was fit to the data, and a F-test was run to assess if the slope was significantly different from zero. Weights were the total number of genes expressed at each stage. Bonferroni correction was applied to correct for multiple testing, considering the 6 ratios tested with mouse EST data (Figure 4A–C; Figure 6; Figure 7; Figure S2): $\alpha = 0.05/6 = 0.0083$. To test for an hourglass-like model, we adjusted a parabola (polynomial model of order 2), as in Hazkani-Covo et al. [3]. We used an ANOVA to estimate if the increase in fit to the data (r) between the linear and parabola models was significant. The same Bonferroni correction was applied to the ANOVA. This test was never significant, providing no evidence for a maximum or a minimum of the ratio during development (Dataset S2).

### Zebrafish In Situ Data

In situ hybridization expression data from ZFIN [16] were retrieved using BGEE [51]. We considered only stages with more than 1000 genes expressed, starting when maternal genes are largely degraded (6 hours post-fertilization [48]). We retrieved all genes with at least one report of expression by in situ hybridization, at each time point of zebrafish development. From this set we extracted two groups (for example essential and non-annotated genes), and analyzed their ratio across development using the same methodology as with ESTs (see above).

### Rate of Protein Evolution

The orthology relationships, and the values of $d_N$ (rate of non-synonymous substitution per codon) and $d_S$ (rate of synonymous substitution per codon) were obtained from Ensembl version 48 [46]. We retrieved zebrafish genes with one-to-one orthologs in Tetraodon nigroviridis and Takifugu rubripes (divergence time is ~32 MYA between the two pufferfish species and ~150 MYA with Danio rerio [53]). We downloaded the pairwise $d_N$ and $d_S$ between Tetraodon and Takifugu, calculated with codeml from the PAML package in the Ensembl pipeline (model = 0, NSsites = 0) [54]. Ensembl considers that $d_S$ values are saturated when they reach a threshold which is 2*median($d_S$). See http://www.ensembl.org/info/about/docs/compara/homology_method.html for further details.

We selected a set of 4937 genes having $d_N$, $d_S$ and Affymetrix expression data. Among them 620 genes were strict singletons in fishes (loss of duplicates after 2R and after the fish-specific genome duplication). At each time point we performed the Spearman correlation between the $d_N/d_S$ ratio and expression, following Davis et al. [1]. A t-statistic was used to assess if the correlation coefficient was different from 0.

For the analysis in mouse we retrieved pairwise $d_N$ and $d_S$ between human and mouse, for genes with one-to-one human orthologs (14,333 genes). We kept only the singletons for 2R genome duplication and separated the 25% with the highest $d_N/d_S$ and the 25% with the lowest $d_N/d_S$ (607 genes in each group). We then compared the expression across development of these two groups using EST data. Using the 10% highest and lowest $d_N/d_S$ gave similar results (not shown).

### Genotypes and Phenotypes

**Zebrafish mutants.** Data on zebrafish mutants were retrieved from the Zebrafish Information Network (http://zfin.org/zf_info/downloads.html, April 2008) [16]. We selected mutant genotypes having a lethal or abnormal phenotype from the file "phenotype.txt", paying attention that they were grown in normal conditions (ZDB-EXP-041102-1). These genotypes were mapped to ZFIN gene IDs using the file "genotype_features.txt" and then to Affymetrix probesets using Biomart [55]. This resulted in a dataset of 252 ZFIN IDs associated with a lethal phenotype (79 Affymetrix probesets), and 2870 ZFIN IDs associated with an abnormal phenotype (461 probesets). Annotated normal phenotype data are rare in ZFIN, due to a lack of report of such mutants in the literature, so we used non-annotated as a reference (7246 ZFIN gene IDs with expression data).

To be sure that the technique used in the phenotype screen did not bias our analysis, we separated the dataset of genotypes having an abnormal phenotype by technique (file "genotype_features.txt"): inversion, transgenic insertion, deficiency, point mutation, translocation, insertion, sequence variant or unspecified. Only transgenic insertions, point mutations and sequence variants provide enough data, with 343, 221 and 2424 ZFIN IDs respectively, corresponding to 309, 171 and 88 Affymetrix probesets respectively (Text S1 and Dataset S1).

**Zebrafish morpholinos.** The morpholinos knock-down phenotypes were downloaded from ZFIN (http://zfin.org/zf_info/downloads.html, April 2008) [16]. We selected morpholinos (file "pheno_environment.txt") giving lethal or abnormal phenotypes (file "phenotype.txt"), paying attention that the genotypes were wild type (file "wildtypes.txt"). The probes were mapped to ZFIN gene IDs using the file "Morpholinos.txt" and then to Affymetrix probesets using Biomart [55]. Only "abnormal" phenotypes provided enough data, with 601 ZFIN IDs corresponding to 256 Affymetrix probesets (Text S1 and Dataset S1).

**Mouse knock-outs.** Data on mouse mutants were retrieved from the Mouse Genome Database (ftp://ftp.informatics.jax.org/

pub/reports/index.html, April 2008) [34]. We extracted from the file MRK_Ensembl_Pheno.rpt all mutant genotypes having an annotated lethal (lethality-embryonic/perinatal, MP:0005374 and lethality-postnatal, MP:0005373), abnormal (other phenotypes detected) or normal phenotype (no phenotype detected, MP:0002873), and their mapping to Ensembl genes. We filtered on the technique used and kept only the mutants obtained with a targeted knock-out. Because different investigators do not report the same phenotypes for the same genes, we removed from the analysis all genes annotated to more than one group. We obtained 50 essential Ensembl genes (lethal phenotype), 164 non essential (normal phenotype), and 1939 whose loss of function is annotated abnormal (Dataset S2). Including genes annotated to more than one group, the group sizes were 1659, 564 and 3721 respectively, and the results were similar (not shown).

### Identification of Duplicate Genes

Gene families were obtained from the HomolEns database version 3 (http://pbil.univ-lyon1.fr/databases/homolens.html), which is based on Ensembl release 41 [46]. HomolEns is build on the same model as Hovergen [56], with genes organized in families, which include pre-calculated alignments and phylogenies. In HomolEns version 3, alignments are computed with MUSCLE [57] (with default parameters), and phylogenetic trees with PhyML [58]. Phylogenies are computed on conserved blocks of the alignments selected with GBLOCKS [59]. Using the TreePattern functionality of the FamFetch client for HomolEns, which allows scanning for gene tree topologies [60], we selected sets of genes with or without duplications on specific branches of the vertebrate phylogenetic tree.

Regarding the fish-specific whole genome duplication, we found 1772 Ensembl IDs for duplicates in zebrafish, 8821 for singletons in zebrafish, 755 mouse orthologs of these duplicates, and 6843 mouse orthologs of these singletons. For the 2R whole genome duplications, we found 986 duplicates and 1266 singletons in zebrafish, and 2448 duplicates and 2705 singletons in mouse (Datasets S1 and S2).

### Gene Ontology Analysis

Over and under representation of GO terms [25] was tested by means of a Fisher exact test, using the Bioconductor package topGO version 1.8.1 [61]. The reference set was all Ensembl genes mapped to a probeset of the zebrafish Affymetrix chip. The "elim" algorithm of topGO was used, allowing to decorrelate the graph structure of the gene ontology, reducing non-independence problems. A False Discovery Rate correction was applied, and gene ontology categories with a FDR <15% were reported.

### Tools

R was used for statistical analysis and plotting (http://www.R-project.org/) [62], in conjunction with Bioconductor packages (http://www.bioconductor.org/, version 2.2)[45]. To retrieve genomic information we used the BioMart tool [55] or connected to the Ensembl MySQL public database [46].

### Supporting Information

**Figure S1** Expression in zebrafish development of genes according to retention after vertebrate 2R whole genome duplications. Median expression profiles of vertebrate specific 2R duplicates in zebrafish in red dashed line and triangles, and of singletons in black solid line and circles. Legend as in Figure 2.
Found at: doi:10.1371/journal.pgen.1000311.s001 (0.53 MB TIF)

**Figure S2** Variation across mouse development of the ratio of expressed vertebrate 2R singletons, relative to duplicates. Legend as in Figure 4.
Found at: doi:10.1371/journal.pgen.1000311.s002 (0.31 MB TIF)

**Figure S3** Variation across zebrafish development of the Spearman correlation between gene sequence evolution and expression. Only singletons genes (for 2R and fish-specific genome duplications) were considered. We used the ratio of the rate of non-synonymous substitutions on the rate of synonymous substitutions ($d_N/d_S$) as a measure of selective pressure. Correlations below the dashed line are significantly different from 0 (p-value <0.05). The x-axis is in logarithmic scale. A gray box on the x-axis indicates the phylotypic period.
Found at: doi:10.1371/journal.pgen.1000311.s003 (0.41 MB TIF)

**Figure S4** Expression in zebrafish development of genes with abnormal mutant phenotypes. Median expression profiles of zebrafish genes inducing abnormal phenotypes when non functional, for 4 different techniques, compared to non-annotated genes in black solid line and circles. The techniques are: morpholinos in purple dashed-dotted line and squares; transgenic insertions in green dashed line and triangles; point mutations in blue dashed line and diamonds; sequence variants in red dotted line and crosses. Points significantly different from the reference curve (non annotated genes) are filled. See Figure S5 for confidence intervals of the difference with the reference curve. The x-axis is in logarithmic scale. A gray box on the x-axis indicates the phylotypic period.
Found at: doi:10.1371/journal.pgen.1000311.s004 (0.41 MB TIF)

**Figure S5** Significance of the expression difference between zebrafish genes inducing abnormal phenotypes when non functional and non-annotated genes for 4 different techniques. These randomization plots refer to Figure S4. Legend as in Figure 2B.
Found at: doi:10.1371/journal.pgen.1000311.s005 (1.05 MB TIF)

**Table S1** Gene Ontology analysis. The two groups analyzed are the genes experiencing an increase of expression along development (late expression, cluster 4) and the genes experiencing a decrease of expression (early expression, cluster 3) (Figure 3). Molecular Function and Biological process ontologies were analyzed with the "elim" algorithm of the Bioconductor package topGO (see Methods).
Found at: doi:10.1371/journal.pgen.1000311.s006 (0.02 MB PDF)

**Dataset S1** Details and characteristics of zebrafish gene sets used in this study. FSGD: Fish Specific whole Genome Duplication.
Found at: doi:10.1371/journal.pgen.1000311.s007 (3.63 MB XLS)

**Dataset S2** Details and characteristics of mouse gene sets used in this study. FSGD: Fish Specific whole Genome Duplication.
Found at: doi:10.1371/journal.pgen.1000311.s008 (1.51 MB XLS)

**Text S1** Supplementary text.
Found at: doi:10.1371/journal.pgen.1000311.s009 (0.03 MB DOC)

### Acknowledgements

We thank Geisler R, Konantz M, Otto GW, Saric M, and Weiler C for making their microarray data publicly available. We thank Jérôme Goudet, Linda Z. Holland, Liliane Michalik and members of the MRR lab for helpful discussions.

## Author Contributions

Conceived and designed the experiments: JR MRR. Performed the experiments: JR. Analyzed the data: JR MRR. Wrote the paper: JR MRR.

## References

1. Davis JC, Brandman O, Petrov DA (2005) Protein evolution in the context of Drosophila development. J Mol Evol 60: 774–785.
2. Castillo-Davis CI, Hartl DL (2002) Genome Evolution and Developmental Constraint in Caenorhabditis elegans. Mol Biol Evol 19: 728–735.
3. Hazkani-Covo E, Wool D, Graur D (2005) In search of the vertebrate phylotypic stage: a molecular examination of the developmental hourglass model and von Baer's third law. J Exp Zoolog B Mol Dev Evol 304: 150–158.
4. Irie N, Sehara-Fujisawa A (2007) The vertebrate phylotypic stage and an early bilaterian-related stage in mouse embryogenesis defined by genomic information. BMC Biology 5: 1.
5. Hanada K, Shiu S-H, Li W-H (2007) The Nonsynonymous/Synonymous Substitution Rate Ratio versus the Radical/Conservative Replacement Rate Ratio in the Evolution of Mammalian Genes. Mol Biol Evol 24: 2235–2241.
6. Yang J, Li WH (2004) Developmental constraint on gene duplicability in fruit flies and nematodes. Gene 340: 237–240.
7. Holland LZ (2007) Developmental biology: A chordate with a difference. Nature 447: 153–155.
8. von Baer KE (1828) Ueber Entwicklungsgeschichte der Thiere: Beobachtung und Reflexion. Königsberg: Bornträger. 271 p.
9. Haeckel E (1874) Anthropogenie oder Entwickelungsgeschichte des Menschen. Leipzig: Engelmann. 732 p.
10. His W (1874) Unsere Körperform und das physiologische Problem ihrer Entstehung. Leipzig: FCW. Vogel. 224 p.
11. Gould SJ (1977) Ontogeny and phylogeny. Cambridge, Mass.: The Belknap Press of Harvard University Press. 501 p.
12. Duboule D (1994) Temporal colinearity and the phylotypic progression: a basis for the stability of a vertebrate Bauplan and the evolution of morphologies through heterochrony. Dev Suppl. pp 135–142.
13. Raff RA (1996) The shape of life: genes, development, and the evolution of animal form. Chicago; London: University of Chicago Press. 520 p.
14. Sanetra M, Begemann G, Becker MB, Meyer A (2005) Conservation and co-option in developmental programmes: the importance of homology relationships. Front Zool 2: 15.
15. Irmler I, Schmidt K, Starck JM (2004) Developmental variability during early embryonic development of zebra fish, Danio rerio. J Exp Zoolog B Mol Dev Evol 302: 446–457.
16. Sprague J, Bayraktaroglu L, Clements D, Conlin T, Fashena D, et al. (2006) The Zebrafish Information Network: the zebrafish model organism database. Nucleic Acids Res 34: D581–585.
17. Poe S, Wake MH (2004) Quantitative tests of general models for the evolution of development. Am Nat 164: 415–422.
18. Blalock EM, ed (2003) A Beginner's Guide to Microarrays: Springer. 368 p.
19. Jaillon O, Aury J-M, Brunet F, Petit J-L, Stange-Thomann N, et al. (2004) Genome duplication in the teleost fish Tetraodon nigroviridis reveals the early vertebrate proto-karyotype. Nature 431: 946–957.
20. Brunet FG, Crollius HR, Paris M, Aury JM, Gibert P, et al. (2006) Gene loss and evolutionary rates following whole-genome duplication in teleost fishes. Mol Biol Evol 23: 1808–1816.
21. Putnam NH, Butts T, Ferrier DEK, Furlong RF, Hellsten U, et al. (2008) The amphioxus genome and the evolution of the chordate karyotype. Nature 453: 1064–1071.
22. Chain FJ, Ilieva D, Evans BJ (2008) Duplicate gene evolution and expression in the wake of vertebrate allopolyploidization. BMC Evol Biol 8: 43.
23. Drummond DA, Bloom JD, Adami C, Wilke CO, Arnold FH (2005) Why highly expressed proteins evolve slowly. Proc Natl Acad Sci USA 102: 14338–14343.
24. Drummond DA, Raval A, Wilke CO (2006) A single determinant dominates the rate of yeast protein evolution. Mol Biol Evol 23: 327–337.
25. Ashburner M, Ball CA, Blake JA, Botstein D, Butler H, et al. (2000) Gene ontology: tool for the unification of biology. The Gene Ontology Consortium. Nat Genet 25: 25–29.
26. Farre D, Bellora N, Mularoni L, Messeguer X, Alba MM (2007) Housekeeping genes tend to show reduced upstream sequence conservation. Genome Biology 8: R140.
27. Poe S (2006) Test of Von Baer's law of the conservation of early development. Evolution 60: 2239–2245.
28. Raff RA (1996) The shape of life : genes, development, and the evolution of animal form. Chicago; London: University of Chicago Press. 520 p.
29. Bininda-Emonds OR, Jeffery JE, Richardson MK (2003) Inverting the hourglass: quantitative evidence against the phylotypic stage in vertebrate development. Proc Biol Sci 270: 341–346.
30. Holland LZ, Gibson-Brown JJ (2003) The Ciona intestinalis genome: when the constraints are off. Bioessays 25: 529–532.
31. Galis F, Metz JA (2001) Testing the vulnerability of the phylotypic stage: on modularity and evolutionary conservation. J Exp Zool 291: 195–204.
32. Wapinski I, Pfeffer A, Friedman N, Regev A (2007) Natural history and evolutionary principles of gene duplication in fungi. Nature 449: 54–61.
33. Richardson MK (1999) Vertebrate evolution: the developmental origins of adult variation. Bioessays 21: 604–613.
34. Eppig JT, Bult CJ, Kadin JA, Richardson JE, Blake JA, et al. (2005) The Mouse Genome Database (MGD): from genes to mice–a community resource for mouse biology. Nucleic Acids Res 33: D471–475.
35. Arthur W (2000) The concept of developmental reprogramming and the quest for an inclusive theory of evolutionary mechanisms. Evol Dev 2: 49–57.
36. Prud'homme B, Gompel N, Carroll SB (2007) Colloquium Papers: Emerging principles of regulatory evolution. Proc Natl Acad Sci USA 104: 8605–8612.
37. Koutsos AC, Blass C, Meister S, Schmidt S, MacCallum RM, et al. (2007) Life cycle transcriptome of the malaria mosquito Anopheles gambiae and comparison with the fruitfly Drosophila melanogaster. Proc Natl Acad Sci USA 104: 11304–11309.
38. Hinman VF, Nguyen AT, Cameron RA, Davidson EH (2003) Developmental gene regulatory network architecture across 500 million years of echinoderm evolution. Proc Natl Acad Sci USA 100: 13356–13361.
39. Nei M (2007) The new mutation theory of phenotypic evolution. Proc Natl Acad Sci USA.
40. Davidson EH, Erwin DH (2006) Gene regulatory networks and the evolution of animal body plans. Science 311: 796–800.
41. Solnica-Krezel L (2005) Conserved patterns of cell movements during vertebrate gastrulation. Curr Biol 15: R213–228.
42. Stern CD, ed (2004) Gastrulation: From Cells to Embryo: Cold Spring Harbor Laboratory Press.
43. Parkinson H, Kapushesky M, Shojatalab M, Abeygunawardena N, Coulson R, et al. (2007) ArrayExpress–a public database of microarray experiments and gene expression profiles. Nucleic Acids Res 35: D747–750.
44. Wu Z, Irizarry RA, Gentleman R, Martinez-Murillo F, Spencer F (2004) A Model-Based Background Adjustment for Oligonucleotide Expression Arrays. Journal of the American Statistical Association 99: 909–917.
45. Gentleman R, Carey V, Bates D, Bolstad B, Dettling M, et al. (2004) Bioconductor: open software development for computational biology and bioinformatics. Genome Biology 5: R80.
46. Hubbard TJ, Aken BL, Beal K, Ballester B, Caccamo M, et al. (2007) Ensembl 2007. Nucleic Acids Res 35: D610–617.
47. Pelegri F (2003) Maternal factors in zebrafish development. Dev Dyn 228: 535–554.
48. Mathavan S, Lee SG, Mak A, Miller LD, Murthy KR, et al. (2005) Transcriptome analysis of zebrafish embryogenesis using microarrays. PLoS Genet 1: 260–276.
49. Schuster EF, Blanc E, Partridge L, Thornton JM (2007) Correcting for sequence biases in present/absent calls. Genome Biol 8: R125.
50. Futschik ME, Carlisle B (2005) Noise-robust soft clustering of gene expression time-course data. J Bioinform Comput Biol 3: 965–988.
51. Bastian F, Parmentier G, Roux J, Moretti S, Laudet V, et al. (2008) Bgee: Integrating and Comparing Heterogeneous Transcriptome Data Among Species. In: Heidelberg SB, ed. Data Integration in the Life Sciences. pp 124–131.
52. Pontius JU, Wagner L, Schuler GD (2003) 21. UniGene: A Unified View of the Transcriptome. In: McEntyre J, Ostell J, eds. The NCBI Handbook. Bethesda, MD: National Library of Medicine (US), NCBI.
53. Benton MJ, Donoghue PC (2007) Paleontological evidence to date the tree of life. Mol Biol Evol 24: 26–53.
54. Yang Z (1997) PAML: a program package for phylogenetic analysis by maximum likelihood. Comput Appl Biosci 13: 555–556.
55. Kasprzyk A, Keefe D, Smedley D, London D, Spooner W, et al. (2004) EnsMart: a generic system for fast and flexible access to biological data. Genome Res 14: 160–169.
56. Duret L, Mouchiroud D, Gouy M (1994) HOVERGEN: a database of homologous vertebrate genes. Nucleic Acids Res 22: 2360–2365.
57. Edgar RC (2004) MUSCLE: multiple sequence alignment with high accuracy and high throughput. Nucleic Acids Res 32: 1792–1797.
58. Guindon S, Gascuel O (2003) A simple, fast, and accurate algorithm to estimate large phylogenies by maximum likelihood. Syst Biol 52: 696–704.
59. Castresana J (2000) Selection of conserved blocks from multiple alignments for their use in phylogenetic analysis. Mol Biol Evol 17: 540–552.
60. Dufayard J-F, Duret L, Penel S, Gouy M, Rechenmann F, et al. (2005) Tree pattern matching in phylogenetic trees: automatic search for orthologs or paralogs in homologous gene sequence databases. Bioinformatics 21: 2596–2603.
61. Alexa A, Rahnenfuhrer J, Lengauer T (2006) Improved scoring of functional groups from gene expression data by decorrelating GO graph structure. Bioinformatics 22: 1600–1607.
62. R Development Core Team (2007) R: A Language and Environment for Statistical Computing. Vienna, Austria: R Foundation for Statistical Computing.